\documentclass[onecolumn,tightenlines,nofootinbib,preprintnumbers,amsmath,amssymb]{revtex4}
\usepackage[usenames,dvipsnames]{color}
\usepackage{caption2}
\usepackage{dcolumn}
\usepackage{graphicx}% Include figure files
\usepackage{subfigure}
\usepackage{epstopdf}
\usepackage{bm}% bold math
\usepackage{lscape}
\usepackage{slashed}
\usepackage{booktabs}
\renewcommand{\arraystretch}{0.9}
\usepackage{setspace}
\usepackage{makecell}

\begin{document}
\begin{spacing}{1.5}
%\nofiles
\title{ The resonance effect for the CP asymmetry associated with the process $\omega \rightarrow \pi^{+}\pi^{-}\pi^{0}$ }
% Force line breaks with \\
\author{Xi-Liang Yuan$^{1}$\footnote{Email: m18937315007@163.com}, Gang L\"{u}$^{1}$\footnote{Email: ganglv66@sina.com}}
\affiliation{\small $^{1}$ Institute of Theoretical Physics, College of Science, Henan University of Technology, Zhengzhou 450001, China}
%\date{\today}
\begin{abstract}

The direct CP asymmetry in the weak decay process of hadrons is commonly attributed to the weak phase of the CKM matrix and the indeterminate strong phase.
We propose a way of creating a strong phase difference between two decay paths involving vector mesons $V= \omega,\rho$ decaying to $\pi^{+}\pi^{-}\pi^{0}$
considering the G-parity suppressed decay process $\rho^{0} \rightarrow \pi^{+}\pi^{-}\pi^{0} $. 
It can lead to a significant CP asymmetry in the interference region. We consider the effects of $\omega$ and $\rho$ mixing on their decays to $\pi^{+}\pi^{-}\pi^{0}$. 
Meanwhile, we calculate the CP asymmetry for the process $B\rightarrow \pi^{+}\pi^{-}\pi^{0}P$ (P refers to pseudoscalar meson) through
 $\omega(\rho) \rightarrow \pi^{+}\pi^{-}\pi^{0}$. 
Besides, we also calculate the integral results of different phase space regions. We hope that our work can help provide some theoretical guidance for future experimental studies of CP asymmetry in these decays.
\end{abstract}
\maketitle

\section{\label{intro}Introduction}
The asymmetry of Charge-Parity (CP) is crucial for our understanding of particle physics and the early universe's evolution. It is a fundamental and significant property of weak interaction and has garnered considerable attention since its initial discovery in 1964 \cite{0JJ1964}. In the Standard Model (SM), CP violation is associated with the weak complex phase in the Cabibbo–Kobayashi–Maskawa (CKM) matrix, which describes the mixing of different generations of quarks. CP asymmetry is a crucial tool to test the standard model (SM). CP asymmetry also probe new physics beyond SM in the field of particle physics. In addition to the weak phase, a substantial strong phase is also necessary for a significant CP asymmetry. Typically, this strong phase is provided by QCDF corrections and phenomenological models. One of the main areas of research in this field is the B meson, which contains a heavy bottom quark and has a large decay rate that makes it ideal for observing CP asymmetry.

The common non-leptonic weak decay of B meson is extensively studied using Perturbative QCD (PQCD) \cite{P2017}, Soft Collinear Effective Theory (SCET) \cite{BB4},  and QCD factorization (QCDF) \cite{N2003}.
The Perturbative QCD method separates the hard part of the process where QCD participates due to transverse momentum and handles it with perturbation theory. It introduces the Sudakov factor to suppress the non-perturbative effects. The non-perturbative contribution is included in the hadron wave function. 
Soft Collinear Effective Theory is the method for calculating hadron matrix elements based on collinear factorization is presented. SCET is a theory that describes the interaction of high-energy quarks with collinear and soft gluons. It has been successfully applied to calculate the decay of B mesons (quark-antiquark bound states involving bottom quarks) as well as the properties of hadron sprays produced in particle collisions where quarks or gluons are generated. Additionally, SCET has been utilized to compute the electroweak interaction of the Higgs boson. SCET possesses the capability to handle multiple soft energy scales and provides a power counting scheme. 
However, in the heavy quark limit within the framework of QCD factorization, for processes involving a quark with a large mass (such as the b quark). Then we can consider taking the mass to infinity and neglecting higher-order contributions proportional to $1/m_b$. The two-body non-leptonic decay amplitude can be expressed as the product of the form factor from the initial state meson to the final state meson and the light cone distribution amplitude of the final state meson in the heavy quark limit. In recent years, the study of B mesons has gained significant prominence as a widely employed method among researchers, owing to the successful resolution of the non-leptonic three-body decay problem. Investigating CP asymmetry in B meson decay processes under resonance effects in QCDF offers a promising avenue for experiment \cite{2lu2021}.

There is a discrepancy between the theoretical predictions and experimental data regarding the CP asymmetry of the $B^{\pm}\rightarrow \omega K^{\pm}$ decay. Theoretical studies using QCD factorization \cite{HY2}, perturbative QCD factorization \cite{HN3} , and Soft Collinear Effective Theory \cite{WW4} have provided values of $0.221_{0.128-0.130}^{+0.137+0.140}$, $0.32_{-0.17}^{+0.15}$ and $0.116_{-0.204-0.011}^{+0.182+0.011}$, respectively. Although these theoretical approaches have large uncertainties, they suggest a significant CP asymmetry in this decay channel. However, experimental measurements by Belle \cite{BB5} and BaBar \cite{BC6} indicate a numerically small CP asymmetry consistent with zero, with the latest world average  being $A_{CP}=-0.02\pm0.04$ \cite{J2020}. At present, these theoretical approaches have not provided a satisfactory solution to this puzzle. 
The investigation of CP asymmetry in multibody final state decay processes associated with $\omega$ mesons represents a significant endeavor, offering valuable insights into this type of CP violation. Experimentally distinguishing between $\omega$ and $\rho$ mesons is an exceedingly challenging task.

We propose a way of creating a strong phase difference between two decay paths involving vector mesons $V=\omega,\rho$ decaying to $\pi^{+}\pi^{-}\pi^{0}$ considering the G-parity suppressed decay process $\rho^{0} \rightarrow \pi^{+}\pi^{-}\pi^{0} $. We will employ the QCDF method to analyze B meson decay and incorporate the resonance effect. This study represents the first attempt to investigate the contribution of the $\omega$-$\rho$ mixing effect to the CP asymmetry of the $\omega \rightarrow \pi^{+}\pi^{-}\pi^{0}$ process involving B-meson decay and the mixing of $\omega$-$\rho$ mesons in the final state. Our analysis indicates that this effect significantly influences the measurement of CP asymmetry in $\omega \rightarrow \pi^{+}\pi^{-}\pi^{0}$ \cite{ZP2017}. These vector mesons of $\omega$ and $\rho$ interference can mix with each other and produce a new strong phase difference between the decay paths. They can also enhance the CP asymmetry in some regions of phase space. At the same time, the vector meson resonances reveal rich information about the particle properties and the interactions among mesons \cite{MDA1979}.

The Experimental differentiation between $\omega$ meson and $\rho$ meson remains unattainable. In this work, we calculate the CP asymmetry from the decay process $\bar{B}^{0}\rightarrow \omega\pi^{0}(\bar K^{0})\rightarrow\pi^{+}\pi^{-}\pi^{0}\pi^{0}(\bar K^{0})$,  $\bar{B}^{0}\rightarrow\omega\eta(\eta^{'})\rightarrow \pi^{+}\pi^{-}\pi^{0}\eta(\eta^{'})$ and $B^{-}\rightarrow \omega\pi^{-}( K^{-})\rightarrow\pi^{+}\pi^{-}\pi^{0}\pi^{-}( K^{-})$ considering the effect of the G-parity suppressed decay process $\rho^{0} \rightarrow \pi^{+}\pi^{-}\pi^{0}$ from QCDF. 
In the second part, we investigate the resonance effect in the four-body decay process by elucidating its physical mechanism in section A, presenting comprehensive computational formalisms for CP asymmetry in section B, and examining the manifestation of localized CP asymmetry in section C.
Subsequently, we comprehensively illustrate the decay processes involved in this study and present a comprehensive analysis of the typical four-body decay amplitude processes in the third section.
In the fourth section, we conduct an analysis of the CP asymmetry results and present the corresponding numerical findings. The outcomes are visually depicted through the use of graphical representations. In addition, calculations are conducted to determine the integral results associated with different phase space regions. The final section encompasses our comprehensive analysis and future prospects pertaining to the research conducted in this paper.

\section{\label{sum}CALCULATION OF CP ASYMMETRY}
\subsection{\label{subsec:form} The introduction of resonance mechanism}
The contribution of vector mesons is added to the photon propagator, that is the photon propagator is mixed with the intermediate $\omega(\rho)$ propagator. According to the vector meson dominance model, $e^{+}e^{-}$ can annihilate into a pair of $\pi^{+}\pi^{-}$, which can then polarize in the vacuum and form vector mesons $\omega(782)$ and $\rho^{0}(770)$. These vector mesons can then decay into $\pi^{+}\pi^{-}$ pairs \cite{NB1967}. The $\omega-\rho$ mixing is due to the difference in quark mass and electromagnetic effects. $\omega(782)$ and $\rho^{0}(770)$ can also be linked by a photon propagator and then result in mixing. Later, in order to study the modification of the intermediate process of photons by the low energy strong interaction, the vector meson-dominant model (VMD) was proposed, which successfully described the interaction between photons and hadronic matter. 
If you use the basis vector of the isospin is represented by $|I,I_3>$ , then the isospin eigenstates can be represented by \cite{6Lu2017}:
\begin{equation}
|\rho_{I} = |1,0>, \quad\quad
|\omega_{I} = |0,0>.	
\end{equation}

We define $|a>$ where $a = \rho, \omega$. The physical states $|a>$ and the isospin states $|a_{I}>$ form a complete set of orthogonal basis vectors, respectively \cite{2lu2022}:
\begin{equation}
I=\sum_a|a><a|=	\sum_{a_I}|a><a|, 
\end{equation}
and
\begin{equation}
\delta=<a|b>=<a_I|b_I>.
\end{equation}

Therefore, the two sets of bases can be converted to each other:
\begin{equation}
|a>=\sum_{b_I}|b_I><b_I|a>, \quad\quad
|a_I>=\sum_{b}|b><b|a_I>.
\end{equation}

Since the mixture of $\omega$ and $\rho$ is relatively small, then the conversion of the above two groups of base vectors can be expressed in the following form:
\begin{equation}
	|\rho>=|\rho_I>-\epsilon|\omega_I>, \quad\quad
	|\omega>=|\omega_I>+\epsilon|\rho_I>,
\end{equation}
where $\epsilon$ represents a small quantity. The matrix form can be expressed as:
\begin{equation}
\left (
\begin{array}{l}
	\rho  
	\\[0.5cm]
	 \omega 
\end{array}
\right )  =
\left (
\begin{array}{lll}
<\rho_{I}|\rho> & \hspace{0.5cm} <\omega_{I}|\rho>  \\[0.5cm]
<\rho_{I}|\omega> &  \hspace{0.5cm}<\omega_{I}|\omega>
\end{array}
\right )
\left (
\begin{array}{l}
	\rho_I  
	\\[0.5cm]
	\omega_I 
\end{array}
\right ) \\
=
\left (
\begin{array}{lll}
 ~~~~1 &\hspace{0.5cm} -F_{\rho\omega}(s) &
\\[0.5cm]
\displaystyle  F_{\rho\omega}(s) &  \hspace{0.5cm}~~~ 1 &
\end{array}
\right ),
\label{L2}
\end{equation}
where $F_{\rho\omega}(s)$ is order $\mathcal{O}(\lambda)$, $(\lambda\ll 1)$ \cite{2lu2022}. The physical state of this transformation can be expressed as:
\begin{eqnarray}
\rho^{0}=\rho^{0}_{I}-F_{\rho\omega}(s)\omega_{I}, &&
\omega=F_{\rho\omega}(s)\rho^{0}_{I}+\omega_{I}.
\label{A}
\end{eqnarray}
In view of representations from the physics and isospin, we make propagator definitions as $D_{V_1V_2}=\left< 0|TV_1V_2|0 \right> $ and $D_{V_1V_2}^{I}=\left< 0|TV_{1}^{I}V_{2}^{I}|0 \right>$, respectively. $V_{1}$ and $V_{2}$ of $D_{V_{1}V_{2}}$ refer to the meson of $\omega$ and $\rho^{0}$. In fact, $D_{V_{1}V_{2}}$ is equal to zero, because there is no two vector meson mixing under the physical representation. Besides, according to the expression for the physical state of the two vector meson mixing, the parameters of $F_{\rho\omega}$
is order of $\mathcal{O}(\lambda)$ ($\lambda\ll 1$). Multiplication of any two or three terms in the equation is of higher order and can be ignored. We represent the inverse propagator in terms of decay width and mass, which are expressed in detail in the following formula \cite{15Li2022}:
\begin{eqnarray}
\Pi_{\rho\omega}=F_{\rho\omega}(s-m^{2}_{\rho}+im_{\rho}\Gamma_{\rho}
	)-F_{\rho\omega}(s-m^{2}_{\omega}+im_{\omega}\Gamma_{\omega}),
\label{A}
\end{eqnarray}
where $\Pi_{\rho\omega}$ is the mixing parameter. $S$ and $m_{V}$ denote the inverse propagator and mass of vector meson $V$ $(V= \omega,\rho)$, respectively. 
$\Gamma_{V}$ $(V= \omega,\rho)$ refers to the decay rate of the vector meson V, we consider the values which come from PDG $\Gamma_{\omega}=89.2\pm0.7\%$ and $\Gamma_{\rho}=1.01^{+0.54}_{-0.36}\pm0.34\times10^{-4}\%$ in our work \cite{PDG,J2003}. Meanwhile, the values of $\omega$ full width is $8.68\pm0.13 $MeV and the values of $\rho$ full width is $149.1\pm0.8 $MeV. 
The propagator $s_{V}$ is associated with the invariant mass $\sqrt s$, and it serves as a robust indicator of CP asymmetry.

The mixing parameters $\omega-\rho$ is extracted from the $e^{+}e^{-}\rightarrow \pi^{+}\pi^{-}$ experimental data \cite{MN2000,P2009}.
In order to better interpret the mixture of $\rho-\omega$ in our work, we define:
\begin{eqnarray}
\widetilde{\Pi}_{\omega \rho}=\frac{(s-m^{2}_{\omega}+im_{\omega}\Gamma_{\omega})\Pi_{\omega \rho}}{(s-m^{2}_{\omega}+im_{\omega}\Gamma_{\omega}-(s-m^{2}_{\rho}+im_{\rho}\Gamma_{\rho}))},
\label{A}
\end{eqnarray}

The mixing parameters of $\widetilde{\Pi}_{\omega \rho}(s)$ is the momentum dependence for $\omega-\rho$ interference  \cite{P2011}. So we can define $\tilde{\Pi}_{V_1V_2}=\frac{s_{V_1}\Pi _{V_1V_2}}{s_{V_1}-s_{V_2}}$ from the above equations, where $\tilde{\Pi}_{ \omega \rho}$ is the mixing parameters for $\omega-\rho$ interferences \cite{MN2000}. These parameters are functions of the momentum and have been measured by Wolfe and Maltman \cite{P2009,P2011}. Based on the following theoretical basis: 
\begin{eqnarray}
\widetilde{\Pi}_{\omega \rho}(s)={\mathfrak{Re}}\widetilde{\Pi}_{\omega \rho}(m_{\omega}^2)+{\mathfrak{Im}}\widetilde{\Pi}_{\omega \rho}(m_{\omega}^2).
\label{A}
\end{eqnarray}
The real and imaginary parts of the $\omega-\rho$ mixing parameter $\widetilde\Pi_{\omega \rho}$ when $s=m^2_{\omega}$ is suited to be:
$\mathfrak{Re}\widetilde{\Pi}_{\omega \rho}(m_{\omega}^2)=-4760\pm440
\rm{MeV}^2$, $
{\mathfrak{Im}}\widetilde{\Pi}_{\omega \rho}(m_{\omega}^2)=-6180\pm3300
\textrm{MeV}^2$.
\subsection{\label{subsec:form}The computation form of CP asymmetry}
In order to vividly explain the decay process of mesons we proposed, we use the $\bar{B}^{0}\rightarrow \omega(\rho)\pi^{0}\rightarrow\pi^{+}\pi^{-}\pi^{0}\pi^{0}$ decay channel as an example to study CP asymmetry. The diagram (a) in Fig.1, $\bar B^{0}$ meson decays into $\pi^0$ and $\pi^{+}\pi^{-}\pi^{0}$ pair which is produced directly by $\rho$ meson. Meanwhile, the middle diagram (b) shows that $\bar B^{0}$ meson decays into $\pi^0$ and $\pi^{+}\pi^{-}\pi^{0}$ pair which is produced directly by $\omega$ meson in Fig.1. However, it is known that $\pi^{+}\pi^{-}\pi^{0}$ pair can also exist by the resonance effect from the intermediate state of $\omega$ or $\rho$ meson. Then we consider the processes of $\rho$ and $\omega$ decay into $\pi^{+}\pi^{-}\pi^{0}$, which are shown in the diagram (c) of Fig.1. 
There are the resonance effects of  $\rho-\omega$ mixing and $\omega-\rho$ mixing respectively. 
Moreover, we have plotted the decay process of $\rho \rightarrow \omega \rightarrow\pi^{+}\pi^{-}\pi^{0}$, which plays a major role in the decay process. But since $\rho \rightarrow\pi^{+}\pi^{-}\pi^{0}$ is depressed
by the G-parity violation in the decay process of $\omega \rightarrow \rho  \rightarrow\pi^{+}\pi^{-}\pi^{0}$, we have neglected the contribution of this process and are not shown in the Fig.1.
\begin{figure}[h]
\centering
\includegraphics[height=4cm,width=16cm]{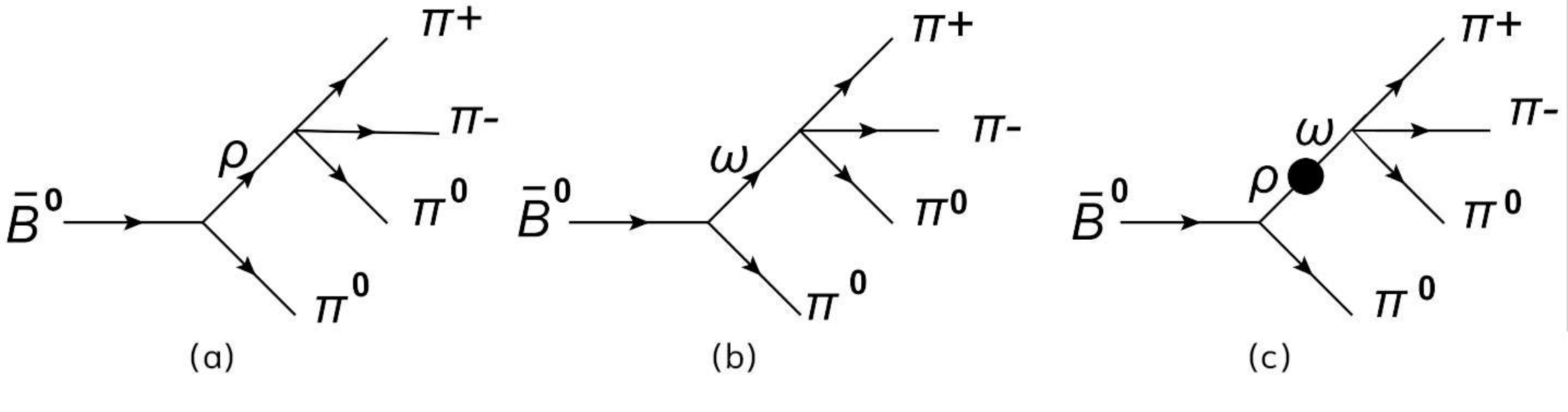}
\caption{The decay diagrams for the channel of $\bar B^{0}\rightarrow \omega(\rho)\pi^{0} \rightarrow \pi^{+}\pi^{-}\pi^{0}\pi^{0}$.}
\label{fig1}
\end{figure}

The interference between mesons not only offers dynamic insights into their properties but also induces significant CP asymmetry from the strong phase
in intermediate vector meson mixing.
We take the decay process of $\bar B^{0}\rightarrow \omega\pi^{0} \rightarrow \pi^{+}\pi^{-}\pi^{0}\pi^{0}$ as an example to study CP asymmetry. We use N to represent the total amplitude used in the calculation, and the physical meaning of N is the sum of the tree contribution($\big<\pi^{+}\pi^{-}\pi^{0}\pi^{0}|H^T|\bar B^{0}\big>$) and Penguin contribution($\big<\pi^{+}\pi^{-}\pi^{0}\pi^{0}|H^P|\bar B^{0}\big>$).

The ratio of the contribution of the penguin diagram to the contribution of the tree diagram produces the phase angle, which affects the CP asymmetry in the decay process. The form is as follows \cite{2lu2021}:
\begin{eqnarray}
N=\big<\pi^{+}\pi^{-}\pi^{0}\pi^{0}|H^T|\bar B^{0}\big>[1+re^{i(\delta+\phi)}],
\label{A'}
\end{eqnarray}
where r is represented by the ratio of the amplitude contribution of the Penguin level to the amplitude contribution of the Tree level in our calculation. The strong phase $\delta$ is generated through the resonance effect in this four body decay process and the weak phase $\phi$ is associated with the CKM matrix. They can affect the CP asymmetry of this decay process.
Furthermore, we can present the detailed formalisms of the tree and penguin amplitude by combining above decay diagrams in Fig.1:

\begin{eqnarray}
	\begin{array}{c}
	\big<\pi^{+}\pi^{-}\pi^{0}\pi^{0}|H^T|\bar B^{0}\big>
	=\frac{\Gamma_{\rho \rightarrow \pi^+ \pi^- \pi^0}t_{\rho}}{s-m^{2}_{\rho}+im_{\rho}\Gamma_{\rho}}
	+\frac{\Gamma_{\omega \rightarrow \pi^+ \pi^- \pi^0}\widetilde{\Pi}_{\omega\rho}t_{\rho}}{(s-m^{2}_{\omega}+im_{\omega}\Gamma_{\omega}) (s-m^{2}_{\rho}+im_{\rho}\Gamma_{\rho})}\\
		\\
	+\frac{\Gamma_{\omega \rightarrow \pi^+ \pi^- \pi^0}t_{\omega}}{s-m^{2}_{\omega}+im_{\omega}\Gamma_{\omega}}
	+\frac{\Gamma_{\rho \rightarrow \pi^+ \pi^- \pi^0}\widetilde{\Pi}_{\omega\rho}t_{\omega}}{(s-m^{2}_{\rho}+im_{\rho}\Gamma_{\rho})(s-m^{2}_{\omega}+im_{\omega}\Gamma_{\omega})},
	\end{array}
\end{eqnarray}

\begin{eqnarray}
	\begin{array}{c}
		\big<\pi^{+}\pi^{-}\pi^{0}\pi^{0}|H^P|\bar B^{0}\big>=\frac{\Gamma_{\rho \rightarrow \pi^+ \pi^- \pi^0}p_{\rho}}{s-m^{2}_{\rho}+im_{\rho}\Gamma_{\rho}}
		+\frac{\Gamma_{\omega \rightarrow \pi^+ \pi^- \pi^0}\widetilde{\Pi}_{\omega\rho}p_{\rho}}{(s-m^{2}_{\omega}+im_{\omega}\Gamma_{\omega}) (s-m^{2}_{\rho}+im_{\rho}\Gamma_{\rho})}\\
		\\
		+\frac{\Gamma_{\omega \rightarrow \pi^+ \pi^- \pi^0}p_{\omega}}{s-m^{2}_{\omega}+im_{\omega}\Gamma_{\omega}}
		+\frac{\Gamma_{\rho \rightarrow \pi^+ \pi^- \pi^0}\widetilde{\Pi}_{\omega\rho}p_{\omega}}{(s-m^{2}_{\rho}+im_{\rho}\Gamma_{\rho})(s-m^{2}_{\omega}+im_{\omega}\Gamma_{\omega})},
	\end{array}
\end{eqnarray}
where $t_{\rho}(p_{\rho})$ and $t_{\omega}(p_{\omega})$ are the tree (penguin) contribution for $\bar B^{0}\rightarrow\rho^{0}\pi^{0}$ and $\bar B^{0}\rightarrow\omega \pi^{0}$ decay processes, respectively. $s$ is the invariant mass squared of
mesons $\pi^+\pi^-\pi^0$ \cite{epjc}, Besieds, we use $\Gamma_{\omega}=\Gamma_{\omega \rightarrow \pi^+ \pi^- \pi^0}$ and $\Gamma_{\rho}=\Gamma_{\rho \rightarrow \pi^+ \pi^- \pi^0}$. Then, we can obtain:
\begin{eqnarray}
re^{i\delta}e^{i\phi} \equiv \frac{	\big<\pi^{+}\pi^{-}\pi^{0}\pi^{0}|H^P|\bar B^{0}\big>}
{	\big<\pi^{+}\pi^{-}\pi^{0}\pi^{0}|H^T|\bar B^{0}\big>},
\quad r_{1}
e^{i(\delta_\lambda+\phi)}\equiv \frac{p_{\omega}}{t_{\rho}},
\quad r_{2} e^{i\delta_\alpha}\equiv \frac{t_{\omega}}{t_{\rho}} ,
\quad r_{3} e^{i\delta_\beta}\equiv \frac{p_{\rho}}{p_{\omega}}, \label{def}
\end{eqnarray}
$\delta_\lambda$, $\delta_\alpha$ and $\delta_\beta$ are strong phases. 
Subsequently, we can substitute the obtained results into Eqs (12) and (13) and perform simplifications:
\begin{eqnarray}
\begin{split}
re^{i\delta}=
& \quad
\frac{r_{1}e^{i\delta_\lambda}r_{3} e^{i\delta_\beta}\Gamma_{\rho}(s-m^{2}_{\omega}+im_{\omega}\Gamma_{\omega})+r_{1}e^{i\delta_\lambda}\Gamma_{\omega}(s-m^{2}_{\rho}+im_{\rho}\Gamma_{\rho})
+r_{1}e^{i\delta_\lambda}\Gamma_{\rho}\widetilde{\Pi}_{\omega\rho}+r_{1}e^{i\delta_\lambda}r_{3} e^{i\delta_\beta}\Gamma_{\omega}\widetilde{\Pi}_{\omega\rho}}
{\Gamma_{\rho}(s-m^{2}_{\omega}+im_{\omega}\Gamma_{\omega})+\Gamma_{\omega}\widetilde{\Pi}_{\omega\rho}+r_{2}e^{i\delta_\alpha}\Gamma_{\rho}\widetilde{\Pi}_{\omega\rho}
+r_{2}e^{i\delta_\alpha}\Gamma_{\omega}(s-m^{2}_{\rho}+im_{\rho}\Gamma_{\rho})},
\end{split}
\end{eqnarray}

The $\phi$ is related to $\frac{V_{ub}V_{ud}^{*}}{V_{tb}V_{td}^{*}}$ from CKM matrix. Therefore, we can get ${\rm sin}\phi =\frac{\eta}{\sqrt{(\rho-\rho^2-\eta^2)^2+\eta^2}}$ and
${\rm cos}\phi = \frac{\rho-\rho^2-\eta^2}{\sqrt{(\rho-\rho^2-\eta^2)^2+\eta^2}}$ for Wolfenstein parameters \cite{MN2000}.
Then we can make the definition about the CP asymmetry:
$A_{cp} = \frac{\left| N \right|^2-\left| \overline{N} \right|^2}{\left| N \right|^2+\left| \overline{N} \right|^2}$.

\subsection{\label{subsec:form}The integral form of CP asymmetry}
In the following analysis, we will focus on the $\omega \rightarrow \pi^+ \pi^- \pi^0$ process. The existing decay width is obtained by fitting the corresponding coupling constant to the partial decay width of the $\omega \rightarrow \pi^+ \pi^- \pi^0$ process (see \cite{F2009} and references therein).
In our work, we consider the resonance effect that generates a new strong phase and influences the CP asymmetry. However, to obtain a more accurate result, we need to integrate the $A_{CP}$ over the whole phase space. The $\omega$-$\rho$ mixing mechanism in $ B\rightarrow \omega(\rho) P$ is distinct from that in previously studied channels like $ B\rightarrow \omega P \rightarrow \rho \pi^0 P \rightarrow \pi^+\pi^-\pi^0P$. 
In previous studies, $\omega$ mesons decayed into $\rho^0$ mesons and then eventually declinated into $\pi^+ \pi^-$ pairs. But in our study, we consider the $\rho^0$ and $\omega$ resonance transformation of $\pi^+ \pi^- \pi^0$ mesons. The decay process has been studied differently due to the distinct properties of $\omega$ and $\rho^0$ mesons \cite{epjc}. We conclude that: 

\begin{equation}
	\frac{d^2 \Gamma_{V\rightarrow123}}{dm^2_{13}m^2_{23}}=\frac{1}{(2\pi)^3}\frac{P}{32M^3}|\Gamma_{V\rightarrow123}|^{2} ,
\end{equation}
where $\Gamma$ is width of decay, P is the phasespace factor $P=-\frac{1}{3}\epsilon_{\mu v \alpha \beta }p^{\mu}p^{v}_{1}p^{\alpha}_{2}\epsilon_{\overline{\mu} \overline{v} \overline{\alpha} \beta }p^{\overline{\mu}}p^{\overline{v}}_{1}p^{\overline{\alpha}}_{2}$. The relation between mass(m) and momentum(p) is that $m^{2}_{ij}=(p_i+p_j)^2, M^2=p^2$, where $p$ denotes the momentum of the final state meson $\pi^+ \pi^- \pi^0$ and  M represents the mass of the Vector meson. We make an equivalent substitution for momentum and mass: $q_1^2=m_{23}^2, q_3^2=m_{23}^2, q_2^2=m_{13}^2=p^2-m^2_{12}-m^2_{23}+p^2_1+p^2_2+p^2_3, p=p_1+p_2+p_3$. Additionally, $m_{12}$ refers to the invariant mass of $\pi^+ \pi^-$, $m_{13}$ pertains to the invariant mass of $\pi^+ \pi^0$ and $m_{23}$ signifies the invariant mass of $\pi^- \pi^0$. The differential decay width of the process $\omega(\rho) \rightarrow \pi^+ \pi^- \pi^0$ is determined by the matrix element:
\begin{eqnarray}
	\begin{array}{c}
	\Gamma_{\omega \rightarrow \pi^+ \pi^- \pi^0}=\frac{M_V h_P h_A}{4f^{3}m_{\omega}}\left[S_{\rho}(q^2_1)(q^2_1+m^2_{\omega})+S_{\rho}(q^2_2)(q^2_2+m^2_{\omega})\right.\\
	\\
	\left.+S_{\rho}(q^2_3)(q^2_3+m^2_{\omega})\right]-\left[\frac{2M_V h_P m^{2}_{\pi} b_A}{f^{3}m_{\omega}}(S_{\rho}(q^2_1)+S_{\rho}(q^2_2)+S_{\rho}(q^2_3))\right] ,
\end{array}
\end{eqnarray}

\begin{eqnarray}
	\begin{array}{c}
		\Gamma_{\rho \rightarrow \pi^+ \pi^- \pi^0}=\frac{M_V h_P h_A}{4f^{3}m_{\rho}}\left[S_{\omega}(q^2_1)(q^2_1+m^2_{\rho})+S_{\omega}(q^2_2)(q^2_2+m^2_{\rho})\right.\\
		\\
		\left.+S_{\omega}(q^2_3)(q^2_3+m^2_{\rho})\right]-\left[\frac{2M_V h_P m^{2}_{\pi} b_A}{f^{3}m_{\rho}}(S_{\omega}(q^2_1)+S_{\omega}(q^2_2)+S_{\omega}(q^2_3))\right] ,
	\end{array}
\end{eqnarray}
where the precise value for $h_P$ emerges from a fit to the electromagnetic pion form factor, the quantity $f$ denotes the pion-decay constant in the chiral limit. $h_A$ and $b_A$ are parameters determined by the radiation decay of vector mesons. By applying the Lagrange equation, we constrain the combination of $m_{V}h_{P}h_{A}/f^3$ for vector meson decay into two pseudoscalar states. The values of $h_P$ fitted from experimental decay widths are consistent within a range of $±10\%$. Furthermore, considering radiation decay in vector mesons, we determine the values of $h_A$ and $b_A$(Unit: GeV) \cite{PDG}. 
\begin{eqnarray}
	\begin{array}{c}
		h_{p}=0.304, \quad	h_{A}=2.1,\\
		\\
		f=0.09, \quad\quad b_{A}=0.27,
	\end{array}
\end{eqnarray}
Additionally, before performing integral operations, we provide a first-order calculation for hadronic three-body decay of vector mesons and estimate uncertainties based on $\omega \rightarrow \pi^+ \pi^- \pi^0$ decay analysis. In our two-body decay analysis, we draw on the values of the coupling constants of decay processes that have been proven to be associated with SU(3) flavor symmetry to deviate from each other by about $\pm 10\%$. The coupling constant $h_P$ also appears in this context, and measurements of particle combinations are present in hadron double-body decay. So we used these data to fit into our analysis. In Eq.(17,18), we use the energy-dependent meson width combined with the meson mass and momentum to define the propagators $S_{\rho}(q^2)$ and $S_{\omega}(q^2)$. The expressions of momenta dependent propagators $S_{\rho}(q^2)$ and $S_{\omega}(q^2)$ are obtained:

\begin{eqnarray}
	\begin{array}{c}
		S_{\rho}(q^2)=\frac{1}{q^2-m^2_{\rho}+i \sqrt{q^2}\Gamma_{\rho}(q^2)}, \quad	S_{\omega}(q^2)=\frac{1}{q^2-m^2_{\omega}+i \sqrt{q^2}\Gamma_{\omega}(q^2)} ,\\
		\\
		\Gamma_{\rho}(q^2)=\Gamma_0 \left[\frac{\sqrt{q^2-4m^2_{\pi}}}{\sqrt{m^2_{\rho}-4m^2_{\pi}}}\right]\frac{m^2_{\rho}}{q^2}, \quad	\Gamma_{\omega}(q^2)=\Gamma_0 \left[\frac{\sqrt{q^2-4m^2_{\pi}}}{\sqrt{m^2_{\omega}-4m^2_{\pi}}}\right]\frac{m^2_{\omega}}{q^2} ,
	\end{array}
\end{eqnarray}
where $\Gamma_0$ denotes the onshell width of mesons. Since $\omega$ meson is usually reconstructed through the decay channel $\omega \rightarrow \pi^+ \pi^- \pi^0$, the calculated CP asymmetry of $B \rightarrow \pi^+ \pi^- \pi^0 P$ with a constant mass near the $\omega$ resonance can be expressed as \cite{EP2013,J2020}:

\begin{equation}
A_{CP}^{\Omega}=\frac{\int^{(m_{\omega+\Delta\omega})^2}_{(m_{\omega-\Delta\omega})^2} \left[\int\left(|\mathcal{N^{-}}|^{2}-|{\mathcal{N^{+}}}|^{2}\right)\mathrm{~d} m_{13}\mathrm{~d} m_{23}\right]\mathrm{~d}m_{123}}{\int^{(m_{\omega+\Delta\omega})^2}_{(m_{\omega-\Delta\omega})^2}\left[\int\left(|\mathcal{N^{-}}|^{2}+|{\mathcal{N^{+}}}|^{2}\right)\mathrm{~d} m_{13}\mathrm{~d} m_{23}\right]\mathrm{~d}m_{123}}.
\end{equation}

The equation involves the decay amplitude of $ B\rightarrow \pi^+\pi^-\pi^0 P$, where $N^{\pm}$ represents it. The squared invariant masses of the systems $m_{\pi^+\pi^0}$, $m_{\pi^-\pi^0}$, and $m_{\pi^+\pi^-\pi^0}$ are denoted by $m_{13}$, $m_{23}$, and $m_{123}$, respectively. In our four-body decay calculation, We will have three the same quality ($m_ {12}$, $m_{13} $, $m_{23}$) equivalent into two the same quality ($m_{13}$, $m_{23}$), and in the vicinity of $\omega$ calculation, Integral limit set in the $(m_{\omega+\Delta\omega})^2$ and $(m_ {\omega-\Delta\omega})^2$. After that we double integrate. Here the range of $m_{\omega+\Delta\omega}$ and $m_{\omega-\Delta\omega}$ values varies within a small range near $\omega$. The mass of the $\omega$ meson is denoted as $m_{\omega}$. The value of $\Delta\omega$ is chosen such that the line shape of $\omega $ is included in the integral interval. When reconstructing the $\omega$ meson experimentally, the cut $\Delta\omega$ is selected to optimize the signal-to-background ratio by considering the convolution for the width of the $\omega$ meson and the momentum resolution of the detector. Typically, $\Delta\omega$ is comparable to the decay width of the $\omega$ meson $\Gamma_{\omega}$. This has been confirmed in previous studies on ALICE collaboration \cite{E2020,HY2021} and we also selected a threshold interval similar to the test in the following calculation.

\section{\label{sec:cpv1}The calculation of decay amplitude}

In the QCD factorization approach, the non-perturbative contribution can be described by the universal distribution amplitude and form factor of mesons, which encompasses both hard and soft contributions. The value of the form factor can be determined through semi-leptonic decay experiments of B mesons or by utilizing the QCD summation method. Moreover, information regarding the light cone distribution amplitude of mesons can also be extracted from other hard scattering partial decay processes. While naive factorization yields leading order decay amplitudes, radiation corrections to all orders including chiral enhancement factors can be calculated in heavy quark limit when neglecting 1/$m_b$ power corrections. To estimate $A_{CP}^{\Omega}$ accurately, we consider CP asymmetry phenomena in multi-body decays are analyzed and studied based on QCDF method. We need to calculate it based on the amplitude of quasi-two-body decay amplitudes. We employ a quasi-two-body decay process to calculate the CP asymmetry, where both tree-level and penguin-level contributions are involved. In the two-body decay of the B-meson, the form factor governing the transition to the final hadron is dominated by non-perturbative effects \cite{P2022}. Hard gluon exchange is the main source of unfactorized contributions of hadronic matrix elements. We used the QCDF method to calculate the amplitudes of the quasi-two-body decays and completed the calculation of the associated hard-scattering kernels. We write all the decay amplitude of $B\rightarrow \omega(\rho)P\rightarrow \pi^{+}\pi^{-}\pi^{0}P$ in QCDF approach using CKM matrix elements of $V_{ub}V^{*}_{ud}$($V_{ub}V^{*}_{us}$) and $V_{tb}V^{*}_{td}$($V_{tb}V^{*}_{ts}$) as:

\begin{eqnarray}
\begin{array}{c}
	N\left(\bar{B}^{0}\rightarrow \omega(\omega \rightarrow \pi^{+}\pi^{-}\pi^{0})  \pi^0 \right) ={\sum_{\lambda}}{\frac{G_FP_{\bar{B}^{0}}\cdot \epsilon ^*\left( \lambda \right)  \cdot \left( p_{\pi ^+}+p_{\pi^-}+p_{\pi^0} \right)\Gamma_{\omega}}{s_{\omega}}}\left\{V_{ub}V_{ud}^{*}\left[\frac{1}{\sqrt{2}}m_{\omega}\epsilon \cdot p_{\pi}f _{\pi}A_{0}^{B\rightarrow \omega}a_2\right.\right.\\
	\\
	\left.-\frac{1}{\sqrt{2}}m_{\omega}\epsilon \cdot p_{\pi}f_{\omega}F_{1}^{B\rightarrow \pi}a_2+\frac{1}{2\sqrt{2}}f_B f_{\pi} f_{\omega}(b_{1}(\omega,\pi)+b_{1}(\pi,\omega))\right] \\
	\\
\left. +V_{tb}V_{td}^{*}\left[\frac{1}{\sqrt{2}}m_{\omega}\epsilon \cdot p_{\pi}f_{\pi}A_{0}^{B\rightarrow \omega}\left( a_4-\frac{1}{2}a_{10}+a_6Q_1-\frac{1}{2}a_{8}Q_1 +\frac{3}{2}a_7-\frac{3}{2}a_9\right)\right.\right.\\
	\\
\left.\left.+\frac{1}{\sqrt{2}}m_{\omega}\epsilon \cdot p_{\pi}f_{\omega}F_{1}^{B\rightarrow \pi}(a_4-\frac{1}{2}a_{10}+2a_3+2a_5+\frac{1}{2}a_7+\frac{1}{2}a_9)-\frac{1}{2\sqrt{2}}f_B f_{\pi} f_{\omega} \right.\right.\\
	\\
\left.\left.(-b_{3}(\omega,\pi)+b_{3}(\pi,\omega)+\frac{1}{2}b_{3}^{e\omega}(\pi,\omega)
+\frac{1}{2}b_{3}^{e\omega}(\omega,\pi)
+\frac{3}{2}b_{4}^{e\omega}(\omega,\pi)+\frac{3}{2}b_{4}^{e\omega}(\pi,\omega))\right]\right\},
\end{array}
\end{eqnarray}

\begin{eqnarray}
\begin{array}{c}
	N\left(\bar{B}^{0}\rightarrow \rho(\rho \rightarrow \pi^{+}\pi^{-}\pi^{0})  \pi^0 \right) ={\sum_{\lambda}}{\frac{G_FP_{\bar{B}^{0}}\cdot \epsilon ^*\left( \lambda \right)  \cdot \left( p_{\pi ^+}+p_{\pi^-}+p_{\pi^0} \right)\Gamma_{\rho}}{s_{\rho}}}\left\{V_{ub}V_{ud}^{*}\left[-\frac{1}{\sqrt{2}}m_{\rho}\epsilon \cdot p_{\pi}f_{\rho}F_{1}^{B\rightarrow \pi}a_2 \right.\right.\\
	\\
	 \left.\left.-\frac{1}{\sqrt{2}}m_{\rho}\epsilon \cdot p_{\pi}f_{\pi}A_{0}^{B\rightarrow \rho}a_2+\frac{1}{2\sqrt{2}}f_B f_{\pi} f_{\rho}(b_{1}(\rho,\pi)+b_{1}(\pi,\rho))\right] \right. \\
	\\
\left. +V_{tb}V_{td}^{*}\left[-\frac{1}{\sqrt{2}}m_{\rho}\epsilon \cdot p_{\pi}f_{\pi}A_{0}^{B\rightarrow \rho}\left( a_4-\frac{1}{2}a_{10}+a_6Q_1-\frac{1}{2}a_{8}Q_1 +\frac{3}{2}a_7-\frac{3}{2}a_9\right)\right.\right.\\
	\\
\left.\left.-\frac{1}{\sqrt{2}}m_{\rho}\epsilon \cdot p_{\pi}f_{\rho}F_{1}^{B\rightarrow \pi}(a_4-\frac{1}{2}a_{10}-\frac{3}{2}a_7-\frac{3}{2}a_9)-\frac{1}{2\sqrt{2}}f_B f_{\pi} f_{\rho} (b_{3}(\rho,\pi)+b_{3}(\pi,\rho)\right.\right.\\
	\\
\left.\left.-\frac{1}{2}b_{3}^{e\omega}(\pi,\rho)
+\frac{1}{2}b_{3}^{e\omega}(\rho,\pi)+2b_{4}(\rho,\pi)+2b_{4}(\pi,\rho)
+\frac{1}{2}b_{4}^{e\omega}(\rho,\pi)+\frac{1}{2}b_{4}^{e\omega}(\pi,\rho))\right]\right\},
\end{array}
\end{eqnarray}

\begin{eqnarray}
	\begin{array}{c}
		N\left(\bar{B}^{0}\rightarrow \omega(\omega \rightarrow \pi^{+}\pi^{-}\pi^{0})  \bar{K}^{0} \right) ={\sum_{\lambda}}{\frac{G_FP_{\bar{B}^{0}}\cdot \epsilon ^*\left( \lambda \right)  \cdot \left( p_{\pi ^+}+p_{\pi^-}+p_{\pi^0} \right)\Gamma_{\omega}}{s_{\omega}}}\left\{V_{ub}V_{us}^{*}m_{\omega}\epsilon \cdot p_{K}f_{\omega}F_{1}^{B\rightarrow K}a_2\right.\\
		\\
		\left. -V_{tb}V_{ts}^{*}\left[m_{\omega}\epsilon \cdot p_{K}f_{K}A_{0}^{B\rightarrow \omega}\left( a_4-\frac{1}{2}a_{10}+a_6Q_2-\frac{1}{2}a_{8}Q_2 \right)\right.\right.\\
		\\
		\left.\left.+m_{\omega}\epsilon \cdot p_{K}f_{\omega}F_{1}^{B\rightarrow K}(2a_3+2a_5+\frac{1}{2}a_7+\frac{1}{2}a_9)+\frac{1}{2}f_B f_{K} f_{\omega}(b_{3}(K,\omega)-\frac{1}{2}b_{3}^{e\omega}(K,\omega))\right]\right\},
	\end{array}
\end{eqnarray}

\begin{eqnarray}
	\begin{array}{c}
		N\left(\bar{B}^{0}\rightarrow \rho(\rho \rightarrow \pi^{+}\pi^{-}\pi^{0})  \bar{K}^{0} \right) ={\sum_{\lambda}}{\frac{G_FP_{\bar{B}^{0}}\cdot \epsilon ^*\left( \lambda \right)  \cdot \left( p_{\pi ^+}+p_{\pi^-}+p_{\pi^0} \right)\Gamma_{\rho}}{s_{\rho}}}\left\{V_{ub}V_{us}^{*}m_{\rho}\epsilon \cdot p_{K}f_{\rho}F_{1}^{B\rightarrow K}a_2 \right.\\
		\\
		\left. -V_{tb}V_{ts}^{*}\left[m_{\rho}\epsilon \cdot p_{K}f_{K}A_{0}^{B\rightarrow \rho}\left( a_4-\frac{1}{2}a_{10}+a_6Q_2-\frac{1}{2}a_{8}Q_2 \right)\right.\right.\\
		\\
		\left.\left.+m_{\rho}\epsilon \cdot p_{K}f_{\rho}F_{1}^{B\rightarrow K}(\frac{3}{2}a_7+\frac{3}{2}a_9)-\frac{1}{2}f_B f_{K} f_{\rho}(b_{3}(K,\rho)-\frac{1}{2}b_{3}^{e\omega}(K,\rho))\right]\right\},
	\end{array}
\end{eqnarray}

\begin{eqnarray}
	\begin{array}{c}
		N\left(\bar{B}^{0}\rightarrow \omega(\omega \rightarrow \pi^{+}\pi^{-}\pi^{0})  \eta^{(')} \right) ={\sum_{\lambda}}{\frac{G_FP_{\bar{B}^{0}}\cdot \epsilon ^*\left( \lambda \right)  \cdot \left( p_{\pi ^+}+p_{\pi^-}+p_{\pi^0} \right)\Gamma_{\omega}}{s_{\omega}}}\left\{V_{ub}V_{ud}^{*}\left[m_{\omega}\epsilon \cdot p_{\eta^{(')}}f _{\omega}F_{1}^{B\rightarrow \eta^{(')}}a_2\right.\right.\\
		\\
		\left.+m_{\omega}\epsilon \cdot p_{\eta^{(')}}f^u_{\eta^{(')}} A_0^{B\rightarrow\omega}a_2+\frac{1}{2}f_B f^u_{\eta^{(')}} f_{\omega}(b_{1}(\omega,\eta^{(')})+b_{1}(\eta^{(')},\omega))\right]\\
		\\
		\left. -V_{tb}V_{td}^{*}\left[m_{\omega}\epsilon \cdot p_{\eta^{(')}}f _{\omega}F_{1}^{B\rightarrow \eta^{(')}}\left( a_4-\frac{1}{2}a_{10}+2a_3+2a_{5} +\frac{1}{2}a_7+\frac{1}{2}a_9\right)\right.\right.\\
		\\
		\left.\left.+m_{\omega}\epsilon \cdot p_{\eta^{(')}}f^u_{\eta^{(')}}A_0^{B\rightarrow\omega}(a_4-\frac{1}{2}a_{10}+2a_3-2a_5+\frac{1}{2}a_9-\frac{1}{2}a_7+a_6Q_3-\frac{1}{2}a_8Q_3-a_6Q_3\frac{f^u_{\eta^{(')}}}{f^s_{\eta^{(')}}} \right.\right.\\
		\\
		\left.\left.+\frac{1}{2}a_8Q_3\frac{f^u_{\eta^{(')}}}{f^s_{\eta^{(')}}}+a_3\frac{f^s_{\eta^{(')}}}{f^u_{\eta^{(')}}}-a_5\frac{f^s_{\eta^{(')}}}{f^u_{\eta^{(')}}}-\frac{1}{2}a_9\frac{f^s_{\eta^{(')}}}{f^u_{\eta^{(')}}}-\frac{1}{2}a_7\frac{f^s_{\eta^{(')}}}{f^u_{\eta^{(')}}} )+\frac{1}{2}f_B f^u_{\eta^{(')}} f_{\omega}(b_{3}(\omega,\eta^{(')})
		+b_{3}(\eta^{(')},\omega),\right.\right.\\
		\\
		\left.\left.-\frac{1}{2}b_{3}^{e\omega}(\eta^{(')},\omega)
		+\frac{1}{2}b_{3}^{e\omega}(\omega,\eta^{(')})+2b_{4}(\omega,\eta^{(')})+2b_{4}(\eta^{(')},\omega)
		+\frac{1}{2}b_{4}^{e\omega}(\omega,\eta^{(')})+\frac{1}{2}b_{4}^{e\omega}(\eta^{(')},\omega))\right]\right\},
	\end{array}
\end{eqnarray}

\begin{eqnarray}
	\begin{array}{c}
		N\left(\bar{B}^{0}\rightarrow \rho(\rho \rightarrow \pi^{+}\pi^{-}\pi^{0})  \eta^{(')} \right) ={\sum_{\lambda}}{\frac{G_FP_{\bar{B}^{0}}\cdot \epsilon ^*\left( \lambda \right)  \cdot \left( p_{\pi ^+}+p_{\pi^-}+p_{\pi^0} \right)\Gamma_{\rho}}{s_{\rho}}}\left\{V_{ub}V_{ud}^{*}\left[m_{\rho}\epsilon \cdot p_{\eta^{(')}}f_{\rho}F_{1}^{B\rightarrow \eta^{(')}}a_2 \right.\right.\\
		\\
		\left.\left.-m_{\rho}\epsilon \cdot p_{\eta^{(')}}f^u_{\eta^{(')}}A_{0}^{B\rightarrow \rho}a_2+\frac{1}{2}f_B f^u_{\eta^{(')}} f_{\rho}(b_{1}(\rho,\eta^{(')})+b_{1}(\eta^{(')},\rho))\right]+V_{tb}V_{td}^{*}\left[m_{\rho}\epsilon \cdot p_{\eta^{(')}}f_{\rho}F_{1}^{B\rightarrow \eta^{(')}} \right. \right.\\
		\\
		\left. \left.\left( a_4-\frac{1}{2}a_{10}-\frac{3}{2}a_7 -\frac{3}{2}a_9\right)+m_{\rho}\epsilon \cdot p_{\eta^{(')}}f^u_{\eta^{(')}}A_{0}^{B\rightarrow \rho}(a_4-\frac{1}{2}a_{10}+2a_3-2a_5+\frac{1}{2}a_9+\frac{1}{2}a_7+a_6Q_3\right.\right.\\
		\\
		\left.\left.-\frac{1}{2}a_8Q_3-a_6Q_3\frac{f^u_{\eta^{(')}}}{f^s_{\eta^{(')}}}+\frac{1}{2}a_8Q_3\frac{f^u_{\eta^{(')}}}{f^s_{\eta^{(')}}}+a_3\frac{f^s_{\eta^{(')}}}{f^u_{\eta^{(')}}}-a_5\frac{f^s_{\eta^{(')}}}{f^u_{\eta^{(')}}}-\frac{1}{2}a_9\frac{f^s_{\eta^{(')}}}{f^u_{\eta^{(')}}}-\frac{1}{2}a_7\frac{f^s_{\eta^{(')}}}{f^u_{\eta^{(')}}} )-\frac{1}{2}f_B f^u_{\eta^{(')}} f_{\rho} \right.\right.\\
		\\
		\left.\left.(-b_{3}(\rho,\eta^{(')})
		+b_{3}(\eta^{(')},\rho)+\frac{1}{2}b_{3}^{e\omega}(\eta^{(')},\rho)
		+\frac{1}{2}b_{3}^{e\omega}(\rho,\eta^{(')})+\frac{3}{2}b_{4}^{e\omega}(\rho,\eta^{(')})+\frac{3}{2}b_{4}^{e\omega}(\eta^{(')},\rho))\right]\right\},
	\end{array}
\end{eqnarray}

\begin{eqnarray}
	\begin{array}{c}
		N\left(B^{-}\rightarrow \omega(\omega \rightarrow \pi^{+}\pi^{-}\pi^{0})  \pi^{-} \right) ={\sum_{\lambda}}{\frac{G_FP_{B^{-}}\cdot \epsilon ^*\left( \lambda \right)  \cdot \left( p_{\pi ^+}+p_{\pi^-}+p_{\pi^0} \right)\Gamma_{\omega}}{s_{\omega}}}\left\{V_{ub}V_{ud}^{*}\left[m_{\omega}\epsilon \cdot p_{\pi}f_{\omega}F_{1}^{B\rightarrow \pi}a_2\right.\right.\\
		\\
		\left.\left. +m_{\omega}\epsilon \cdot p_{\pi}f_{\pi}A_{0}^{B\rightarrow \omega}a_1+m_{\omega}\epsilon \cdot p_{\pi}f_{\pi}A_{0}^{B\rightarrow \omega}a_1+m_{\omega}\epsilon \cdot p_{\pi}f_{\pi}A_{0}^{B\rightarrow \omega}a_1+\frac{1}{2}f_Bf_{\pi}f_{\omega}(b_{2}(\pi,\omega)+b_{2}(\omega,\pi))\right]\right.\\
		\\
		\left.-V_{tb}V_{td}^{*}\left[m_{\omega}\epsilon \cdot p_{\pi}f_{\pi}A_{0}^{B\rightarrow \omega}\left(a_4+a_{10}+a_6Q_4+a_{8}Q_4 \right)+m_{\omega}\epsilon \cdot p_{\pi}f_{\omega}F_{1}^{B\rightarrow \pi}(a_4-\frac{1}{2}a_{10}\right.\right.\\
		\\
		\left.\left.+2a_3+2a_5+\frac{1}{2}a_7+\frac{1}{2}a_9)+\frac{1}{2}f_B f_{\pi} f_{\omega}(b_{3}(\pi,\omega)+b_{3}(\omega,\pi)+b_{3}^{e\omega}(\pi,\omega)+b_{3}^{e\omega}(\omega,\pi))\right]\right\},
	\end{array}
\end{eqnarray}

\begin{eqnarray}
	\begin{array}{c}
		N\left(B^{-}\rightarrow \rho(\rho \rightarrow \pi^{+}\pi^{-}\pi^{0}) \pi^{-} \right) ={\sum_{\lambda}}{\frac{G_FP_{B^{-}}\cdot \epsilon ^*\left( \lambda \right)  \cdot \left( p_{\pi ^+}+p_{\pi^-}+p_{\pi^0} \right)\Gamma_{\rho}}{s_{\rho}}}\left\{V_{ub}V_{ud}^{*}\left[m_{\rho}\epsilon \cdot p_{\pi}f_{\rho}F_{1}^{B\rightarrow \pi}a_2\right.\right.\\
		\\
		\left. \left. +m_{\rho}\epsilon \cdot p_{\pi}f_{\pi}A_{0}^{B\rightarrow \rho}a_1+\frac{1}{2}f_B f_{\pi} f_{\rho}(b_{2}(\pi,\rho)-b_{2}(\rho,\pi))\right]-V_{tb}V_{td}^{*}\left[m_{\rho}\epsilon \cdot p_{\pi}f_{\pi}A_{0}^{B\rightarrow \rho}\left(a_4+a_{10}+a_6Q_4+a_{8}Q_4 \right)\right.\right.\\
		\\
		\left.\left.-m_{\rho}\epsilon \cdot p_{\pi}f_{\rho}F_{1}^{B\rightarrow \pi}(a_4-\frac{1}{2}a_10-\frac{3}{2}a_7-\frac{3}{2}a_9)+\frac{1}{2}f_B f_{\pi} f_{\rho}(b_{3}(\pi,\rho)-b_{3}(\rho,\pi)+b_{3}^{e\omega}(\pi,\rho)-b_{3}^{e\omega}(\rho,\pi))\right]\right\},
	\end{array}
\end{eqnarray}

\begin{eqnarray}
	\begin{array}{c}
		N\left(B^{-}\rightarrow \omega(\omega \rightarrow \pi^{+}\pi^{-}\pi^{0})  K^{-} \right) ={\sum_{\lambda}}{\frac{G_FP_{B^{-}}\cdot \epsilon ^*\left( \lambda \right)  \cdot \left( p_{\pi ^+}+p_{\pi^-}+p_{\pi^0} \right)\Gamma_{\omega}}{s_{\omega}}}\left\{V_{ub}V_{us}^{*}\left[m_{\omega}\epsilon \cdot p_{K}f_{\omega}F_{1}^{B\rightarrow K}a_2\right.\right.\\
		\\
		\left.\left. +m_{\omega}\epsilon \cdot p_{K}f_{K}A_{0}^{B\rightarrow \omega}a_1+\frac{1}{2}f_Bf_Kf_{\omega}b_{2}(K,\omega)\right]-V_{tb}V_{ts}^{*}\left[m_{\omega}\epsilon \cdot p_{K}f_{K}A_{0}^{B\rightarrow \omega}\left( a_4+a_{10}+a_6Q_5\right.\right.\right.\\
		\\
		\left.\left.\left.+a_{8}Q_5 \right)+m_{\omega}(\epsilon \cdot p_{K})f_{\omega}F_{1}^{B\rightarrow K}(2a_3+2a_5+\frac{1}{2}a_7+\frac{1}{2}a_9)+\frac{1}{2}f_B f_{K} f_{\omega}(b_{3}(K,\omega)-b_{3}^{e\omega}(K,\omega))\right]\right\},
	\end{array}
\end{eqnarray}

\begin{eqnarray}
	\begin{array}{c}
		N\left(B^{-}\rightarrow \rho(\rho \rightarrow \pi^{+}\pi^{-}\pi^{0}) K^{-} \right) ={\sum_{\lambda}}{\frac{G_FP_{B^{-}}\cdot \epsilon ^*\left( \lambda \right)  \cdot \left( p_{\pi ^+}+p_{\pi^-}+p_{\pi^0} \right)\Gamma_{\rho}}{s_{\rho}}}	\times \left\{V_{ub}V_{us}^{*}\left[m_{\rho}\epsilon \cdot p_{K}f_{\rho}F_{1}^{B\rightarrow K}a_2\right.\right.\\
		\\
		\left.\left.  +m_{\rho}\epsilon \cdot p_{K}f_{K}A_{0}^{B\rightarrow \rho}a_1+\frac{1}{2}f_B f_{K} f_{\rho}(b_{2}(K,\rho))\right]-V_{tb}V_{ts}^{*}\left[m_{\rho}(\epsilon \cdot p_{K})f_{K}A_{0}^{B\rightarrow \rho}\left( a_4+a_{10}\right.\right.\right.\\
		\\
		\left.\left.\left.+a_6Q_5+a_{8}Q_5 \right)+m_{\rho}(\epsilon \cdot p_{K})f_{\rho}F_{1}^{B\rightarrow K}(\frac{3}{2}a_7+\frac{3}{2}a_9)+\frac{1}{2}f_B f_{K} f_{\rho}(b_{3}(K,\rho)+b_{3}^{e\omega}(K,\rho))\right]\right\},
	\end{array}
\end{eqnarray}
where $f_{\pi}$, $f_{B}$, $f_{\omega(\rho)}$ are the decay constants and $a_{1,2...n}$ are related to the Wilson coefficient $C_{i}$ \cite{PRD2000}. $A_{0}^{B\rightarrow\rho}$, $A_{0}^{B\rightarrow\omega}$ and $F_{1}^{B\rightarrow\pi}$ are the form factor from the non-perturbative contribution. Moreover, $b_{3}$, $b_{3}^{e\omega}$ and $b_{4}^{e\omega}$ are the annihilation contributions \cite{J2016}, $\epsilon$ and $p_{\pi}$ are the polarization vector and momentum of $\pi$ meson, respectively. Otherwise the operator Q, which is introduced to make the calculation easier in our work \cite{PR2000,PB2001}, They are defined as follows:

\begin{eqnarray}
	\begin{array}{c}
		Q1=\frac{-2m_{\pi^{0}}^{2}}{(m_b+m_d)(m_d+m_d)}, \quad	Q2=\frac{-2m_{K^{0}}^{2}}{(m_b+m_d)(m_d+m_s)},\quad	Q3=\frac{-2m_{\eta^{(')}}^{2}}{(m_b+m_d)m_{s^{'}}},\\
		\\
		Q4=\frac{-2m_{\pi^{\pm}}^{2}}{(m_b+m_s)(m_u+m_d)}, \quad	Q5=\frac{-2m_{K^{\pm}}^{2}}{(m_b+m_u)(m_u+m_s)}.
	\end{array}
\end{eqnarray}
The mass $m$ in Eq.(31) represents the quark mass associated with the meson component.
We use phenomenological parameters to express the end point integrals of these logarithmic divergence from the hard scattering process of the spectator quark, and we include the twist-3 distribution amplitude of the final pseudo-scalar meson to study the annihilation amplitude \cite{PB2001}. In the annihilation decay process, b is the annihilation coefficient where $b_{1,2}$,
$b_{3,4}$ and $b_{3,4}^{e\omega}$ correspond to the effective operator $Q_{1,2}$, QCD penguin operator $Q_{3-6}$ and the weak electric penguin operator $Q_{7-10}$, respectively \cite{10M2003}. The amplitude is parameterized by the contributions of tree-level and penguin-level, where we introduce the parameters $X_H$ and $X_A$. We define $X_{H}=\int_{0}^{1} \frac{d{y}}{1-y}$ and $X_{A}=\int_{0}^{1} \frac{d{x}}{x}$ to handle the annihilation contribution, then we take the asymptotic form of the amplitude of the meson distribution and bring it into the solution which can not be ignored \cite{J2015}. We consider both the hard scattering of the spectator quark and the annihilation contribution, which give us information about the strong phase. The values of some parameters in the amplitude are as follows:
\begin{table}[!ht]
	\renewcommand
	\arraystretch {3}
	\centering %
	\renewcommand{\arraystretch}{3.0} %
	\caption{Inside Parameter Value}
	\setlength{\tabcolsep}{5mm}
	\begin{tabular}{|c|c|c|c|}
		\hline
		$f_{\omega}=0.195\pm0.005$	& {\makecell[c]{$f_{\rho}=0.216\pm0.003$}}
		
		&{\makecell[c]{$f_{\pi}=0.131\pm0.009$}}
		& $f_{K}=0.155\pm0.002$
		\\ \hline  $f_{B}=0.18\pm0.04$     & $f^u_{\eta^{(')}}=(1.07\pm0.02)f_\pi $ & $f^s_{\eta^{(')}}=(1.34\pm0.06)f_\pi $ & $F_{1}^{B\rightarrow \pi}=0.28\pm0.05$
		\\ \hline  $F_{1}^{B\rightarrow K}=0.31\pm0.02$ & $F_{1}^{B\rightarrow \eta^{(')}}=0.132\pm0.003$ & $A_{0}^{B\rightarrow \rho}=0.37\pm0.06$ & $A_{0}^{B\rightarrow \omega}=0.281\pm0.04$
		\\ \hline  
	\end{tabular}
\end{table}

\section{\label{sum}Numerical results}
\subsection{\label{subsec:form}Diagram of calculation results}
\begin{figure}[!htbp]
	\centering
	\begin{minipage}[h]{0.3\textwidth}
		\centering
		\includegraphics[height=4.5cm,width=6cm]{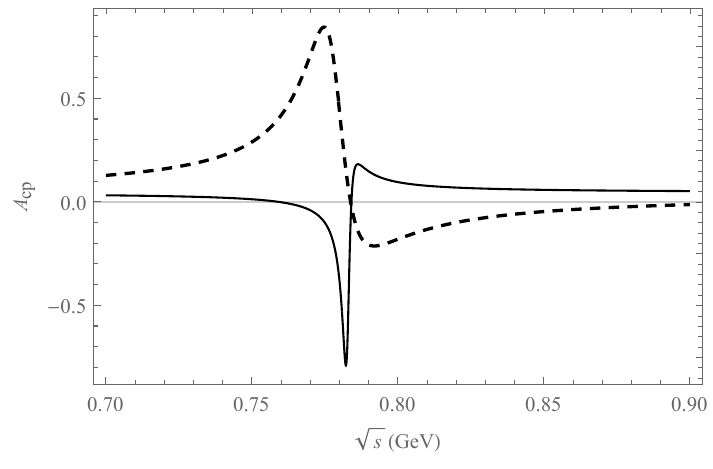}
		\caption{Plot of $A_{CP}$ as a function of $\sqrt{s}$, the solid line is the decay channel of $\bar B^{0}\rightarrow \pi^{+}\pi^{-}\pi^{0}\pi^{0}$, the half-dotted line is the decay channel of $\bar B^{0}\rightarrow \pi^{+}\pi^{-}\pi^{0}\bar K^{0}$.}
		\label{fig2}
	\end{minipage}
	\quad
	\begin{minipage}[h]{0.3\textwidth}
		\centering
		\includegraphics[height=4.5cm,width=6cm]{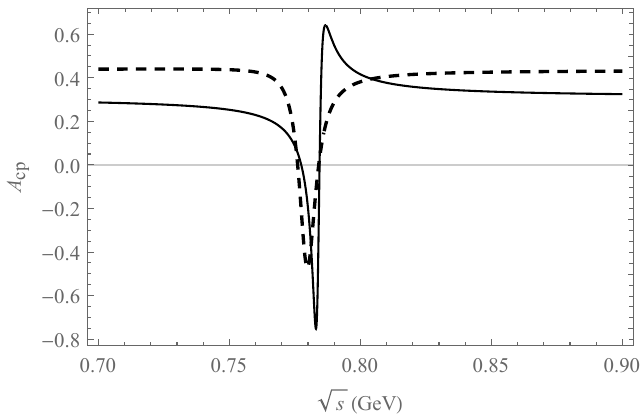}
		\caption{Plot of $A_{CP}$ as a function of $\sqrt{s}$, the solid line is the decay channel of $\bar B^{0}\rightarrow \pi^{+}\pi^{-}\pi^{0}\eta$, the half-dotted line is the decay channel of $\bar B^{0}\rightarrow \pi^{+}\pi^{-}\pi^{0}\eta^{'}$.}
		\label{fig3}
	\end{minipage}
	\quad
	\begin{minipage}[h]{0.3\textwidth}
		\centering
		\includegraphics[height=4.5cm,width=6cm]{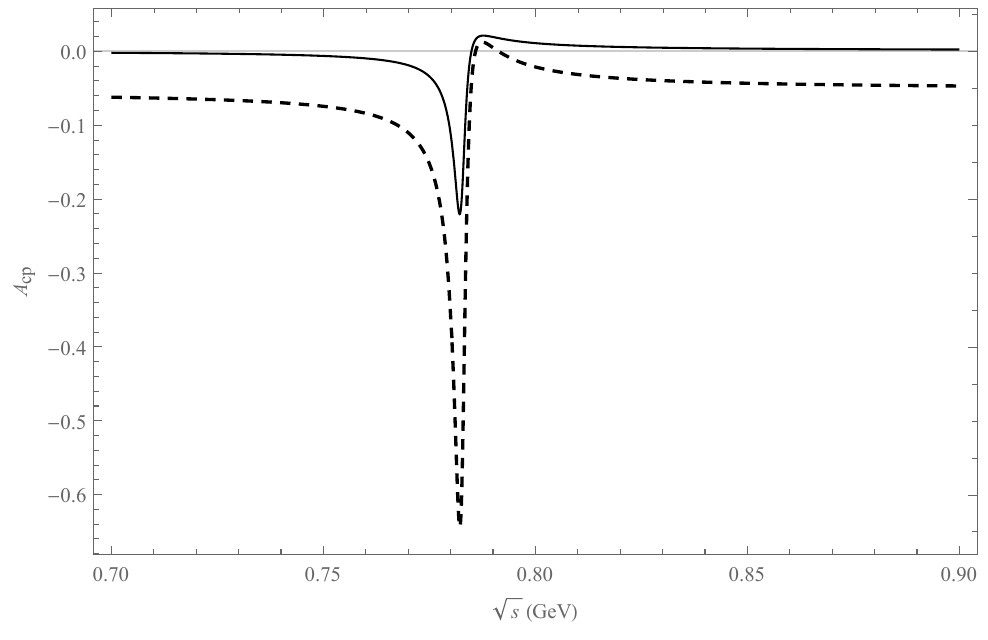}
		\caption{Plot of $A_{CP}$ as a function of $\sqrt{s}$, the solid line is the decay channel of $ B^{-}\rightarrow \pi^{+}\pi^{-}\pi^{0}\pi^{-}$, the half-dotted line is the decay channel of $ B^{-}\rightarrow \pi^{+}\pi^{-}\pi^{0}K^{-}$.}
		\label{fig4}
	\end{minipage}
\end{figure}

We thoroughly consider the previous experimental findings and perform calculations on the resonance effect of vector meson mixing. Consequently, we present the results depicted in Fig. 2 to Fig.10, which illustrate the interrelation between CP asymmetry (r and strong phase $\delta$) and $\sqrt{s}$.
Based on the Particle Data Group (PDG) data, the masses of $\omega$ and $\rho$ are approximately estimated to be 0.782 GeV and 0.770 GeV, respectively \cite{PDG}. To visualize the resonance effect around mass $\omega$, we choose the region 0.70-0.90 GeV under our theoretical framework, which is the main region of resonance for the decay process $\omega(\rho)\rightarrow \pi^{+}\pi^{-}\pi^{0}$  to construct the figures. 
The invariant mass of $\pi^{+}\pi^{-}\pi^{0}$ is displayed in the vicinity of the $\omega$ meson's mass, thus utilizing a range of 0.70 GeV-0.90 GeV for the figures and this enables observation of the overall CP asymmetry \cite{35b2008}. We see that the CP asymmetry changes drastically for the $\bar{B}^{0}\rightarrow \pi^{+}\pi^{-}\pi^{0}\pi^{0}(\bar K^{0}, \eta,\eta^{'})$ and $B^{-}\rightarrow \omega\pi^{-}(K^{-})\rightarrow\pi^{+}\pi^{-}\pi^{0}\pi^{-}( K^{-})$ decay modes due to the $\rho\rightarrow \pi^{+}\pi^{-}\pi^{0}$ and $\omega\rightarrow \pi^{+}\pi^{-}\pi^{0}$ resonances in Fig.\ref{fig2}, Fig.\ref{fig3} and Fig.\ref{fig4} where $\omega$ dominates.

For the $\bar{B}^{0}\rightarrow \pi^{+}\pi^{-}\pi^{0}\pi^{0}$ decay mode, significant CP asymmetries ranging from $-78.5\%$ to $18.2\%$ are observed in the resonance regions of $\omega$ and $\rho$, as shown in Fig.\ref{fig2}. In the case of $\bar{B}^{0}\rightarrow \pi^{+}\pi^{-}\pi^{0}\bar K^{0}$ decay, large CP asymmetries ranging from $84.2\%$ to $-21.9\%$ are found in the same resonance regions of $\omega$ and $\rho$, as depicted in Fig.\ref{fig2}. Furthermore, it is noteworthy that the range of variation for CP asymmetry is similar between the two decay modes: $\bar{B}^{0}\rightarrow \pi^{+}\pi^{-}\pi^{0}\bar K^ {0} $ and $\bar{B}^{0}\rightarrow \pi^{+}\pi^{-}\pi^{0}\pi^{0}$.

We observe CP asymmetries in the decay channel $\bar B^{0}\rightarrow \pi^{+}\pi^{-}\pi^{0}\eta$ ranging from $-74.7\%$ to $64.1\%$ within the resonance range of $\omega$ and $\rho$, as shown in Fig.\ref{fig3}. In the case of $\bar B^{0}\rightarrow \pi^{+}\pi^{-}\pi^{0}\eta'$, we find significant CP asymmetries ranging from $47.1\%$ to $43.5\%$ within the same resonance region depicted in Fig.\ref{fig3}. The decay channel $\bar B^{0}\rightarrow \pi^{+}\pi^{-}\pi^{0}\eta$ exhibits larger CP asymmetry variations compared to that of $\bar B^{0}\rightarrow \pi^{+}\pi^{-}\pi^{0}\eta^{'}$ within the mass resonances of $\omega$($\rho$).

On the other hand, we get the CP asymmetries ranging from $-21.8\%$ to $2.26\%$ at the $\omega$ and $\rho$ resonance range for the decay channel of $B^{-}\rightarrow \pi^{+}\pi^{-}\pi^{0}\pi^{-}$
in Fig.\ref{fig4}. The CP asymmetries range from $-64.5\%$ to $1.30\%$ in the decay channel of $B^{-}\rightarrow \pi^{+}\pi^{-}\pi^{0}K^{-}$ which also has a large change in the CP asymmetry in the resonance of $\omega$ and $\rho$ mass.
We can see that the resonance effect is mainly around the $\omega$ and $\rho$ mass range for all the decay channels.

\begin{figure}[!htbp]
\centering
\begin{minipage}[h]{0.3\textwidth}
\centering
\includegraphics[height=4.5cm,width=6cm]{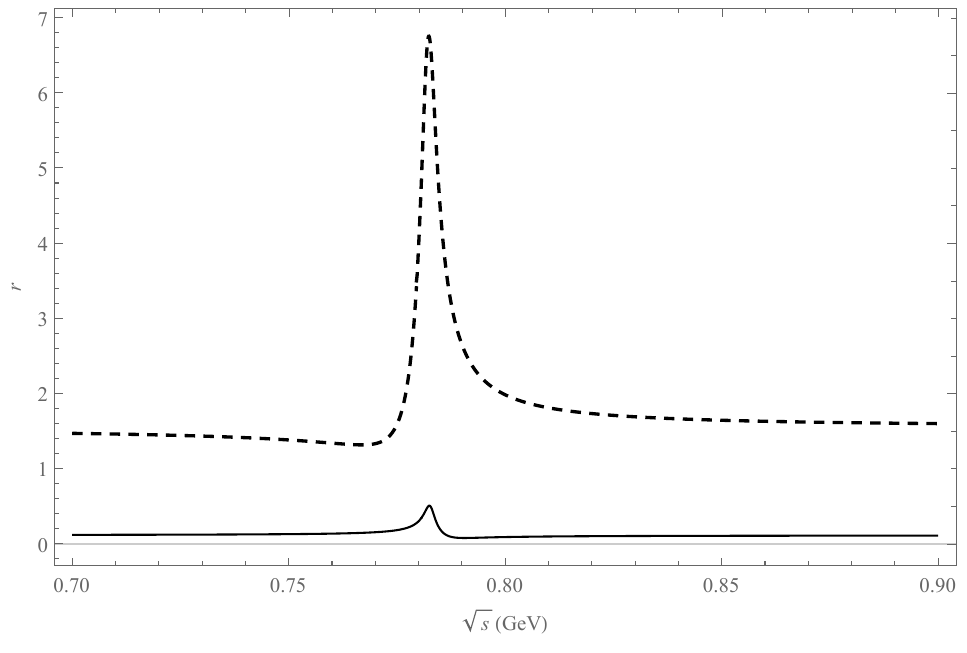}
\caption{Plot of $r$ as a function of $\sqrt{s}$, the solid line is the decay channel of $\bar B^{0}\rightarrow \pi^{+}\pi^{-}\pi^{0}\pi^{0}$, the half-dotted lineis the decay channel of $\bar B^{0}\rightarrow \pi^{+}\pi^{-}\pi^{0}\bar K^{0}$.}
\label{fig5}
\end{minipage}
\quad
\begin{minipage}[h]{0.3\textwidth}
\centering
\includegraphics[height=4.5cm,width=6cm]{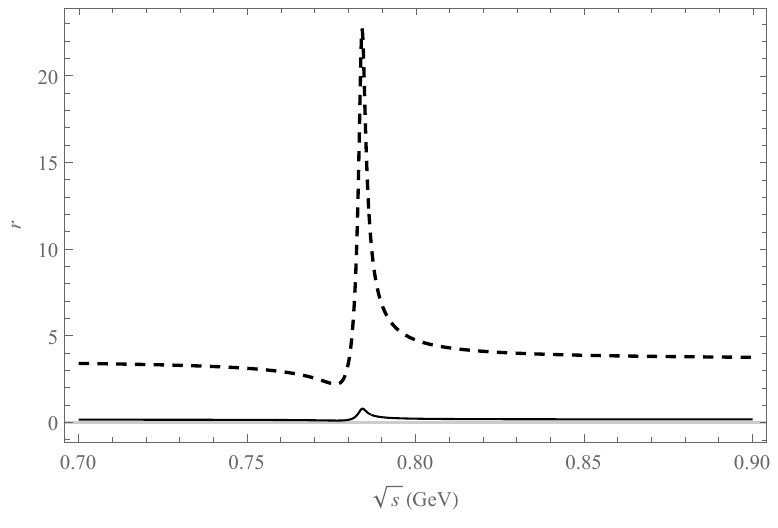}
\caption{Plot of $r$ as a function of $\sqrt{s}$, the solid line is the decay channel of $\bar B^{0}\rightarrow \pi^{+}\pi^{-}\pi^{0}\eta$, the half-dotted lineis the decay channel of $\bar B^{0}\rightarrow \pi^{+}\pi^{-}\pi^{0}\eta^{'}$.}
\label{fig6}
\end{minipage}
\quad
\begin{minipage}[h]{0.3\textwidth}
	\centering
	\includegraphics[height=4.5cm,width=6cm]{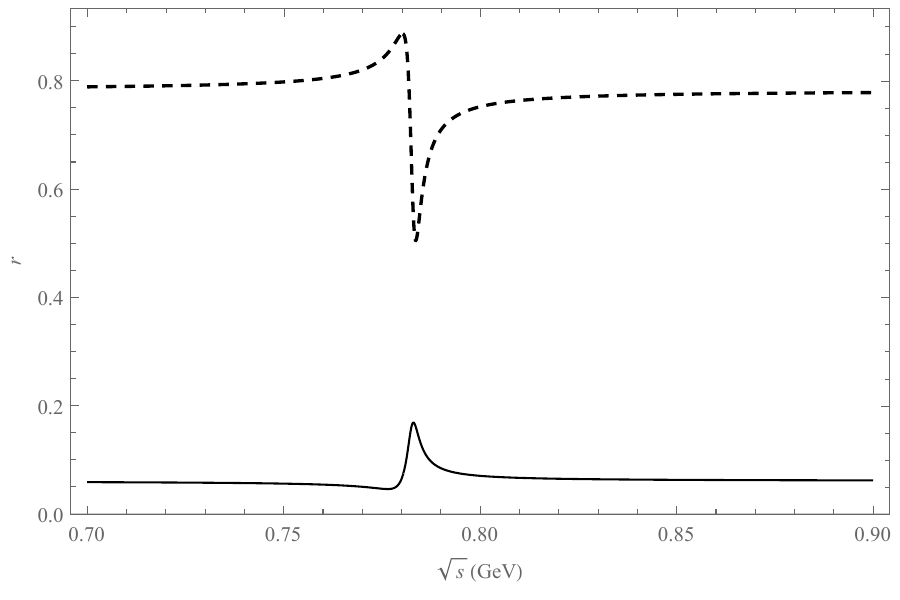}
	\caption{Plot of $r$ as a function of $\sqrt{s}$, the solid line is the decay channel of $ B^{-}\rightarrow \pi^{+}\pi^{-}\pi^{0}\pi^{-}$, the half-dotted lineis the decay channel of $ B^{-}\rightarrow \pi^{+}\pi^{-}\pi^{0}K^{-}$.}
	\label{fig7}
\end{minipage}
\end{figure}

The ratio r of the penguin-level contribution and tree-level contribution influences the CP asymmetry from Eq.(14). We show the relation between r and $\sqrt{s}$ using the central parameter values of CKM matrix elements in Fig.\ref{fig5}, Fig.\ref{fig6} and Fig.\ref{fig7}. The rangeability of r in the $\omega$ and $\rho$ resonance range is bigger for the decay mode of $B\rightarrow \pi^{+}\pi^{-}\pi^{0}(\eta,\pi^-)$. We also notice that r varies are not significant at the $\omega$ mass region from the decay modes of $B\rightarrow \pi^{+}\pi^{-} \bar K^{0}(\eta^{'},K^-)$.

\begin{figure}[!htbp]
\centering
\begin{minipage}[h]{0.3\textwidth}
\centering
\includegraphics[height=4.5cm,width=6cm]{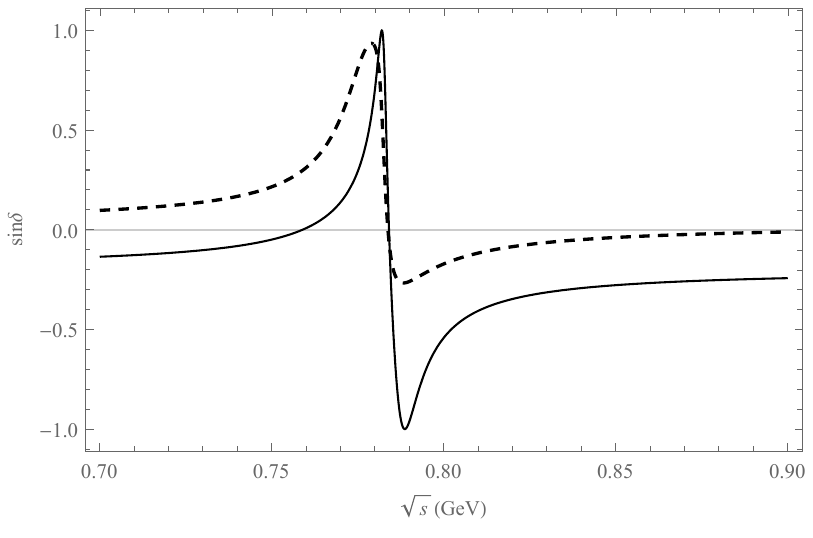}
\caption{Plot of ${\rm{sin}}\delta$ as a function of $\sqrt{s}$, the solid line is the decay channel of $\bar B^{0}\rightarrow \pi^{+}\pi^{-}\pi^{0}\pi^{0}$, the half-dotted lineis the decay channel of $\bar B^{0}\rightarrow \pi^{+}\pi^{-}\pi^{0}\bar K^{0}$.}
\label{fig8}
\end{minipage}
\quad
\begin{minipage}[h]{0.3\textwidth}
\centering
\includegraphics[height=4.5cm,width=6cm]{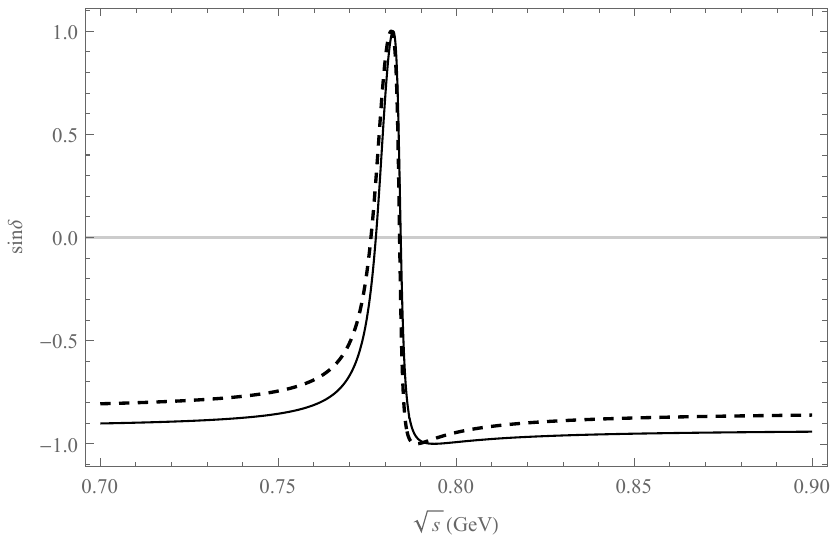}
\caption{Plot of ${\rm{sin}}\delta$ as a function of $\sqrt{s}$, the solid line is the decay channel of $\bar B^{0}\rightarrow \pi^{+}\pi^{-}\pi^{0}\eta$, the half-dotted lineis the decay channel of $\bar B^{0}\rightarrow \pi^{+}\pi^{-}\pi^{0}\eta^{'}$.}
\label{fig9}
\end{minipage}
\quad
\begin{minipage}[h]{0.3\textwidth}
	\centering
	\includegraphics[height=4.5cm,width=6cm]{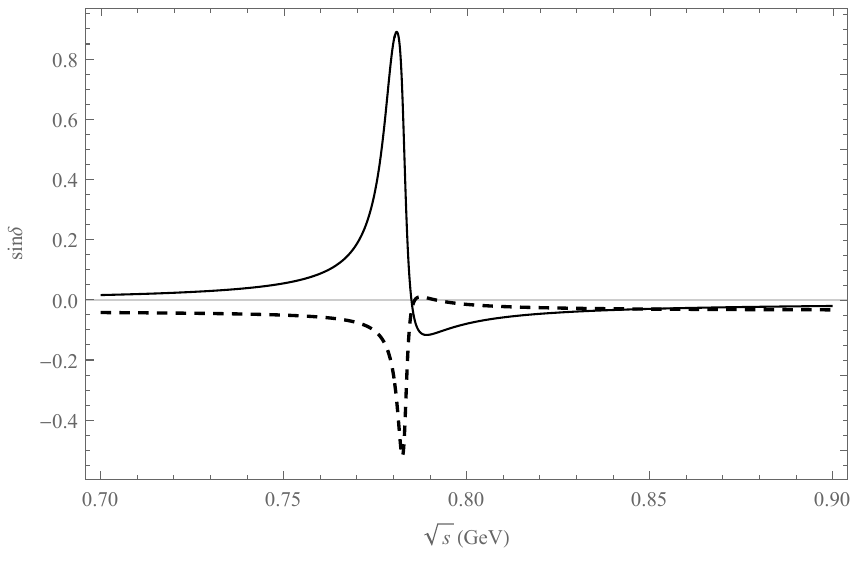}
	\caption{Plot of ${\rm{sin}}\delta$ as a function of $\sqrt{s}$, the solid line is the decay channel of $ B^{-}\rightarrow \pi^{+}\pi^{-}\pi^{0}\pi^{-}$, the half-dotted lineis the decay channel of $ B^{-}\rightarrow \pi^{+}\pi^{-}\pi^{0}K^{-}$.}
	\label{fig10}
\end{minipage}
\end{figure}

We show the plots of ${\rm{sin}}\delta$ versus $\sqrt{s}$ from the central values of the CKM matrix elements that are related to the weak phases Fig.\ref{fig8}, Fig.\ref{fig9} and Fig.\ref{fig10}, to illustrate how the CP asymmetry depends on the strong phase from Eq.(15). In Fig.\ref{fig8}, we can clearly see the variation of ${\rm{sin}}\delta$ in the $\omega$ and $\rho$ resonance regions for the $\bar B^{0}\rightarrow \pi^{+}\pi^{-}\pi^{0}\pi^{0}(\bar K^0)$ decay mode, which suggests that the strong phases play a major role in these processes. Similarly, in Fig.\ref{fig9} and Fig.\ref{fig10}, we observe that ${\rm{sin}}\delta$ changes significantly in the $\omega$ and $\rho$ resonance regions. We also notice that ${\rm{sin}}\delta$ has a large variation near the $\omega$ region regardless for all decay modes and it is not affected by the final meson.

\subsection{\label{subsec:form}Value of localised CP asymmetry}
The CP asymmetry measurement in B meson decay has become more precise due to the large amount of data collected by experiments in recent years and we mentioned that in our previous content \cite{36Lu2019}. The invariant differential cross section of inclusive $\omega(782)$ meson production at midrapidity ($|y|<0.5$) in pp collisions at $\sqrt{s}=7$ TeV was measured utilizing the ALICE detector at the Large Hadron Collider across a transverse momentum range of $2 < p_T < 17$ GeV/c.
The $\omega$ meson was reconstructed through its decay to $\pi^{+}\pi^{-}\pi^{0}$, where the $\pi^{0}$ subsequently decays into two photons. This approach necessitated the reconstruction of charged tracks in the ALICE central tracking system, which comprises the inner tracking system (ITS) and the time projection Chamber (TPC), as well as the reconstruction of photons using both the electromagnetic calorimeter (EMCal) and the photon spectrometer (PHOS). Additionally, photons were reconstructed utilizing the photon conversion method (PCM), which takes advantage of the central tracking system's ability to reconstruct photons from electron–positron track pairs. A comprehensive description of the ALICE detector system and its performance can be found in Refs. \cite{ALI2,ALI3,ALI4}, respectively.

Combined with the experimental observation interval, we select the energy range of CP asymmetry producing peak value in the image, and integrate the CP asymmetry results in this region. In our work, we are the first to calculate the CP asymmetry for the $\bar B^{0}\rightarrow \omega \bar K^{0}(\pi^{0}) \rightarrow \pi^{+}\pi^{-}\pi^{0}\bar K^{0}(\pi^{0})$, $\bar B^{0}\rightarrow \omega \eta(\eta^{'}) \rightarrow \pi^{+}\pi^{-}\pi^{0}\eta(\eta^{'})$ and $B^{-}\rightarrow \pi^{+}\pi^{-}\pi^{0}K^{-}(\pi^-)$ decay modes by integrating over the invariant masses of $m_{\pi^{+}\pi^{-}\pi^{0}}$ in the range of $0.75$ GeV-$0.82$ GeV from the $\rho$ and $\omega$ resonance regions since the mass of $\rho$ is $0.770$ GeV and the mass of $\omega$ is $0.782$ GeV \cite{37plb2021}. The results and the rate of change are given in Table I. We find that the resonance effect clearly affects the CP asymmetry results for the decay processes of $B\rightarrow \pi^{+}\pi^{-}\pi^{0}P$. 

\begin{table}[!ht]
	\renewcommand
	\arraystretch {3}
	\centering %
	\renewcommand{\arraystretch}{3.0} %
	\caption{The comparison of $A^\Omega _{cp}$ from $\omega-\rho$ mixing with $\omega \rightarrow \pi^{+}\pi^{-}\pi^{0}$ }
	\setlength{\tabcolsep}{5mm}
	\begin{tabular}{|c|c|c|c|}
		\hline
		Decay channel
		& {\makecell[c]{Without $\omega-\rho$ mixing\\(0.75-0.82 \textrm{GeV})}}
		
		&{\makecell[c]{$\omega-\rho$ mixing\\(0.75-0.82 \textrm{GeV})}}
		& Rate of change
		\\ \hline  $\bar B^{0}\rightarrow \omega(\rho) \pi^{0} \rightarrow \pi^{+}\pi^{-}\pi^{0}\pi^{0}$     & 0.095$\pm$0.003$\pm$0.005 & -0.015$\pm$0.010$\pm$0.012 & $84.21\%$
		\\ \hline  $\bar B^{0}\rightarrow \omega(\rho) \bar K^{0} \rightarrow \pi^{+}\pi^{-}\pi^{0}\bar K^{0}$   & -0.140$\pm$0.012$\pm$0.011   & 0.107$\pm$0.032$\pm$0.041 & $23.57\%$
		\\ \hline  $\bar B^{0}\rightarrow \omega(\rho) \eta \rightarrow \pi^{+}\pi^{-}\pi^{0}\eta$           & 0.274$\pm$0.001$\pm$0.003  &0.257$\pm$0.028$\pm$0.008 & $6.20\%$
		\\ \hline  $\bar B^{0}\rightarrow       \omega(\rho) \eta^{'} \rightarrow \pi^{+}\pi^{-}\pi^{0}\eta^{'}$    & 0.405$\pm$0.020$\pm$0.027  &0.160$\pm$0.110$\pm$0.087 & $60.49\%$
		\\ \hline  $B^{-}\rightarrow \omega(\rho) \pi^{-} \rightarrow \pi^{+}\pi^{-}\pi^{0}\pi^{-}$           & 0.015$\pm$0.003$\pm$0.004  &-0.012$\pm$0.003$\pm$0.009 & $20.01\%$
		\\ \hline  $B^{-}\rightarrow \omega(\rho)K^{-} \rightarrow \pi^{+}\pi^{-}\pi^{0}K^{-}$            & -0.089$\pm$0.008$\pm$0.012  &-0.010$\pm$0.009$\pm$0.011 & $88.76\%$
		\\ \hline
	\end{tabular}
\end{table}

As indicated in Table I, we present the decay channels calculated along with their corresponding CP asymmetry integral values. The CP asymmetry is particularly prominent near 0.782 GeV, with the integration result being the highest. In addition, we also provide the rate of change of the local integral in the range of 0.75-0.82 GeV, which is convenient to see the influence of resonance effect on CP asymmetry \cite{38C2020}. The mixing results have two uncertainties under the resonance mechanism in our work: the first one is due to the decay width and Wolfstein parameters, and the second one is due to the inherent limitations of the QCDF method, there is a certain error in the processing of numerical results.

\section{\label{sum}SUMMARY AND CONCLUSION}
We demonstrate that CP asymmetry is significantly increased by the resonance effect of $V\rightarrow \pi^{+}\pi^{-}\pi^{0}$ $(V=\rho, \omega)$ from the $B \rightarrow \pi^{+}\pi^{-}\pi^{0}\pi (K,\eta, \eta^{'})$ decay modes when the invariant mass of $\pi^{+}\pi^{-}\pi^{0}$ is close to the $\rho$ and $\omega$ resonance regions in QCD factorization. The interference of $\omega-\rho$ generate new strong phases that affect the CP asymmetry from the $ B \rightarrow \pi^{+}\pi^{-}\pi^{0}\pi(K,\eta^{'})$ decay modes. The resonance has a great impact on the CP asymmetry under the mixing of the $\omega$ and $\rho$. 
For the localized CP asymmetry associated with $\omega(\rho) \rightarrow \pi^{+}\pi^{-}\pi^{0}$, we calculate the integral results obtained from specific phase space regions. The localized CP asymmetry undergoes a significant change due to the resonances of $\omega$ and $\rho$, which arise from the G-parity suppressed decay process of $\rho^{0} \rightarrow \pi^{+}\pi^{-}\pi^{0}$. To observe the impact of the mixing mechanism, we also calculate the results without mixing at the same time as the phase space integration. The resonance effect has a great influence for the CP asymmetry generation in the same energy range. In particular, the decay processes such as $\bar B^{0}\rightarrow \omega(\rho) \pi^{0} \rightarrow \pi^{+}\pi^{-}\pi^{0}\pi^{0}$ and $B^{-}\rightarrow \omega(\rho) K^{-} \rightarrow \pi^{+}\pi^{-}\pi^{0}K^{-}$, the CP asymmetry with mixing mechanism increases by $80\%$ compared with the CP asymmetry without mixing mechanism.

We calculate the CP asymmetry in the decay process by using the $\omega$ energy variation interval similar to the experimental observation. By considering the photon conversion process and accounting for systematic errors inherent in the experiment, we compared and analyzed our mixed and non-mixed results to discern discrepancies between each decay process. The CP asymmetry of these three decay processes $\bar{B}^{0}\rightarrow \omega(\rho) \rightarrow \pi^{+}\pi^{-}\pi^{0}\eta(\eta^{(')},\bar{K}^0)$, can reach $25.7\%$($16\%$, $10.7\%$), which are larger than the asymmetry produced by the decay processes of $\bar{B}^{0}\rightarrow \omega(\rho) \rightarrow \pi^{+}\pi^{-}\pi^{0}\pi^{0}$ and $B^{-}\rightarrow \omega(\rho) \rightarrow \pi^{+}\pi^{-}\pi^{0}\pi^{-}(K^{-})$. Large decay channels resulting in CP asymmetry can be better observed experimentally. The experimental approach involves initially detecting the final state meson of $\pi^{+}\pi^{-}\pi^{0}$ followed by reconstructing the decay channel to identify the intermediate state particle. Our aim is to use the resonance effect to compute at the same threshold interval as in the experiment, providing new insights into CP asymmetry research. We can obtain conclusive outcomes that will facilitate enhanced analysis of experimental measurements. We hope this work will offer a good theoretical basis for experimental exploration and lead to a bright future of particle physics.

\section{Acknowledgments}
This work was supported by  Natural Science Foundation of Henan (Project No.232300420115).

%\newpage

\end{spacing}
\end{document}